\documentclass[twocolumn]{aastex631}
\newcommand{\LCO}{\affiliation{Las Cumbres Observatory, 6740 Cortona Drive, Suite 102, Goleta, CA 93117-5575, USA}}
\newcommand{\UCSB}{\affiliation{Department of Physics, University of California, Santa Barbara, CA 93106-9530, USA}}

\newcommand{\UCD}{\affiliation{Department of Physics and Astronomy, University of California, Davis, 1 Shields Avenue, Davis, CA 95616-5270, USA}}

\newcommand{\UA}{\affiliation{Steward Observatory, University of Arizona, 933 North Cherry Avenue, Tucson, AZ 85721-0065, USA}}

\newcommand{\JHU}{\affiliation{Department of Physics and Astronomy, The Johns Hopkins University, 3400 North Charles Street, Baltimore, MD 21218, USA}}

\newcommand{\GeminiNorth}{\affiliation{Gemini Observatory, 670 North A`ohoku Place, Hilo, HI 96720-2700, USA}}

\newcommand{\Catalyst}{\altaffiliation{LSSTC Catalyst Fellow}}
\newcommand{\USzeged}{\affiliation{Department of Experimental Physics, Institute of Physics, University of Szeged, D\'om t\'er 9, Szeged, 6720 Hungary}}
\newcommand{\ELKHSZTE}{\affiliation{ELKH-SZTE Stellar Astrophysics Research Group, H-6500 Baja, Szegedi \'ut, Kt. 766, Hungary}}
\newcommand{\Konkoly}{\affiliation{Konkoly Observatory, Research Centre for Astronomy and Earth Sciences, E\"otv\"os Lor\'and Research Network (ELKH), Konkoly-Thege Mikl\'os \'ut 15-17, 1121 Budapest, Hungary}}

\submitjournal{The Astrophysical Journal Letters}
\received{2022 October 12}
\revised{2022 November 18}
\accepted{2022 November 20}

\begin{document}

\title{JWST Imaging of the Cartwheel Galaxy Reveals Dust Associated with SN~2021afdx}

\correspondingauthor{Griffin Hosseinzadeh}
\email{griffin0@arizona.edu}

\author[0000-0002-0832-2974]{Griffin~Hosseinzadeh}
\UA
\author[0000-0003-4102-380X]{David J.\ Sand}
\UA
\author[0000-0001-5754-4007]{Jacob E.\ Jencson}
\JHU
\author[0000-0003-0123-0062]{Jennifer E.\ Andrews}
\GeminiNorth
\author[0000-0003-4702-7561]{Irene Shivaei}
\UA
\author[0000-0002-4924-444X]{K.\ Azalee Bostroem}
\Catalyst\UA
\author[0000-0001-8818-0795]{Stefano Valenti}
\UCD

\author[0000-0003-4610-1117]{Tam\'as Szalai}
\USzeged\ELKHSZTE\Konkoly

\author[0000-0003-0035-6659]{Jamison Burke}
\LCO\UCSB
\author[0000-0003-4253-656X]{D.\ Andrew Howell}
\LCO\UCSB
\author[0000-0001-5807-7893]{Curtis McCully}
\LCO\UCSB
\author[0000-0001-9570-0584]{Megan Newsome}
\LCO\UCSB
\author[0000-0003-0209-9246]{Estefania Padilla Gonzalez}
\LCO\UCSB
\author[0000-0002-7472-1279]{Craig Pellegrino}
\LCO\UCSB
\author[0000-0003-0794-5982]{Giacomo Terreran}
\LCO\UCSB

\begin{abstract}

We present near- and mid-infrared (0.9--18 \micron) photometry of supernova (SN) 2021afdx, which was imaged serendipitously with the James Webb Space Telescope (JWST) as part of its Early Release Observations of the Cartwheel Galaxy. Our ground-based optical observations show it is likely to be a Type~IIb SN, the explosion of a yellow supergiant, and its infrared spectral energy distribution (SED) ${\approx}200$~days after explosion shows two distinct components, which we attribute to hot ejecta and warm dust. By fitting models of dust emission to the SED, we derive a dust mass of $(3.8_{-0.3}^{+0.5}) \times 10^{-3}\ M_\odot$, which is the highest yet observed in a Type~IIb SN but consistent with other Type~II SNe observed by the Spitzer Space Telescope. We also find that the radius of the dust is significantly larger than the radius of the ejecta, as derived from spectroscopic velocities during the photospheric phase, which implies that we are seeing an infrared echo off of preexisting dust in the progenitor environment, rather than dust newly formed by the SN. Our results show the power of JWST to address questions of dust formation in SNe, and therefore the presence of dust in the early universe, with much larger samples than have been previously possible.

\end{abstract}


\section{Introduction} \label{sec:intro}

Radio observations of luminous quasars in the early universe (redshift $z \gtrsim 6$, age $\lesssim 1$~Gyr) show them to be large dust reservoirs (${\gtrsim}10^8\ M_\odot$; \citealt{bertoldi_high-excitation_2003,gall_genesis_2011,hashimoto_big_2019}; though see \citealt{bakx_alma_2020} for a recent counterexample). With progenitor lifetimes of only tens of Myr, dust condensation in the expanding ejecta of core-collapse supernovae (SNe) has been proposed as the major source of dust in these early galaxies (see \citealt{gall_production_2011} for a review). This would require the production of up to $1\ M_\odot$ of dust per SN, with ejecta models predicting it would condense during the first 1--2 yr after explosion (e.g., \citealt{dwek_evolution_2007,dwek_evolution_2019}, but see \citealt{wesson_observational_2021} and \citealt{niculescu-duvaz_dust_2022} for alternate discussions). However, nebular observations of SNe in the local universe have for the most part not directly confirmed these large dust masses. Compilations of near- and mid-infrared (IR) observations of SNe yield warm dust masses in the range of $10^{-6}{-}10^{-2}\ M_\odot$ \citep{szalai_twelve_2013,tinyanont_systematic_2016,szalai_comprehensive_2019}.

When the search for SN dust extends to even older SNe, or into the far-IR or radio wavelengths, there is less of a discrepancy between the required and observed dust masses. For example, unambiguous evidence of $0.4{-}0.7\ M_\odot$ of newly formed cold dust has been confirmed in SN~1987A through observations with Herschel and the Atacama Large Millimeter/submillimeter Array \citep{matsuura_herschel_2011,indebetouw_dust_2014}. Additionally, large masses of cold dust have been detected in much older Galactic SN remnants \citep{barlow_herschel_2010,gomez_cool_2012,arendt_interstellar_2014,lau_old_2015,de_looze_dust_2017,temim_massive_2017}, suggesting significant amounts of dust formation occurring in the decades after explosion.

Thermal dust emission typically peaks in the mid-IR (5--10 \micron), putting it out of reach of ground-based observations. Until now, the state of the art in observations of dust in extragalactic SNe was using the Spitzer Space Telescope before it ran out of cryogen in 2009 (the Spitzer Cold Mission) to observe the full near-IR to mid-IR spectral energy distribution (SED) of 12 SNe~II (\citealt{kotak_earlytime_2005,kotak_spitzer_2006,kotak_dust_2009,meikle_spitzer_2007,andrews_photometric_2011,meikle_dust_2011,szalai_dust_2011}; see \citealt{szalai_twelve_2013} and \citealt{priestley_constraining_2020} for the full sample). Since then, Spitzer has observed dozens more SNe out to 4.5~\micron{} during its Warm Mission \citep{fox_disentangling_2010,fox_spitzer_2011,tinyanont_systematic_2016,szalai_comprehensive_2019}, but in this wavelength regime, the thermal and line emission from the SN ejecta can dominate over any dust emission, so it is difficult to measure dust properties and masses.

An additional complication is the fact that dust masses cannot be fully constrained when the dust is optically thick \citep[e.g.,][]{meikle_spitzer_2007}. A large amount of dust can be hidden behind an optically thick surface layer without changing the observed SED. This happens at dust masses around $10^{-3}\ M_\odot$ \citep{meikle_spitzer_2007}, although this number also depends on the dust radius and composition. Therefore it is possible that many of the previous dust measurements are in fact lower limits. However, in cases where contemporaneous observations at shorter wavelengths exist, it can be difficult or impossible to reconcile large amounts of optically thick dust with an unextinguished optical SED \citep[e.g.,][]{priestley_constraining_2020,wesson_observational_2021}.

\begin{figure*}[p]
    \centering
    \includegraphics[width=\textwidth]{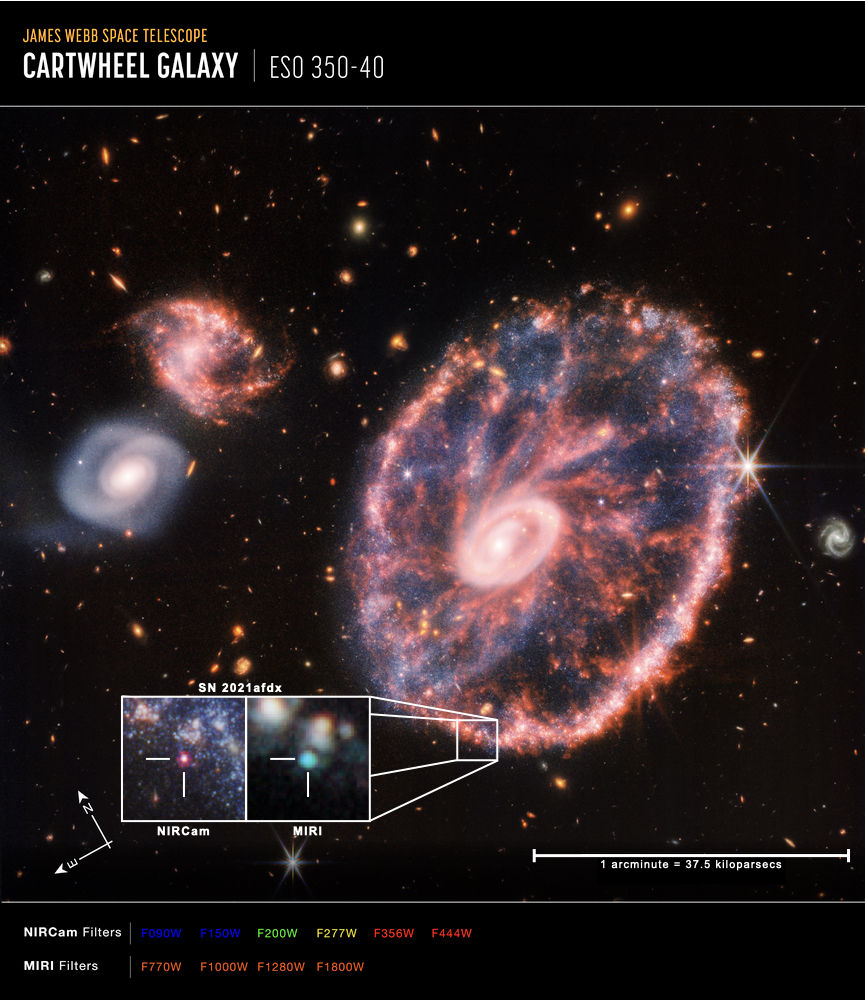}
    \caption{Composite image of the Cartwheel Galaxy taken with JWST's NIRCam and MIRI instruments, with $7\farcs6 \times 7\farcs6$ insets of SN~2021afdx. Credit: NASA, ESA, CSA, STScI, Webb ERO Production Team.}
    \label{fig:galaxy}
\end{figure*}

The first images from the James Webb Space Telescope (JWST) were released on 2022 July 12 \citep{pontoppidan_jwst_2022}, reopening our window into the space-based mid-IR. Images of the Cartwheel Galaxy, taken with the Near-Infrared Camera (NIRCam; \citealt{rieke_overview_2005} and the Mid-Infrared Instrument (MIRI; \citealt{rieke_mid-infrared_2015}), were released shortly thereafter, on 2022 August 2 (Figure~\ref{fig:galaxy}). As first noted by \cite{engesser_detection_2022}, SN~2021afdx is detected in these images, taken at phases of 197 (NIRCam) and 200 (MIRI) rest-frame days after the last prediscovery nondetection.

In this Letter, we use these images to construct the full near- to mid-IR SED of SN~2021afdx with the goal of constraining dust formation in its ejecta, the first opportunity to do this type of analysis in the past decade. In Section~\ref{sec:obs_opt}, we describe our supporting ground-based optical observations, and in Section~\ref{sec:obs_ir}, we measure photometry on the space-based IR images. We fit dust models to the resulting SED and compare to previous measurements in Section~\ref{sec:dust}. In Section~\ref{sec:discuss}, we investigate whether the dust was newly formed in the SN ejecta or whether it existed in the progenitor environment before explosion. We conclude by looking forward to the future of SN dust measurements with JWST in Section~\ref{sec:conclude}.

\section{Observations and Data Reduction} \label{sec:obs}

\subsection{Ground-based Optical} \label{sec:obs_opt}

SN~2021afdx was discovered by the Asteroid Terrestrial-impact Last Alert System (ATLAS; \citealt{tonry_atlas:_2018}) on 2021 November 23.308 UT at R.A.\ 00\textsuperscript{h}37\textsuperscript{m}42\fs580 and decl.\ $-33\degr43\arcmin25\farcs28$ \citep{tonry_atlas_2021a}, $32''$ southeast of the center of the Cartwheel Galaxy (ESO 350-40). We downloaded the full survey light curve from the ATLAS Forced Photometry Server. The explosion time is not very well constrained, as the transient only peaked ${\approx}1.5$~mag above the typical limiting magnitude of the survey, so we adopt the last prediscovery nondetection on 2021 November 21.352 UT as $\mathrm{phase} = 0$ throughout our analysis.

We also obtained multiband follow-up photometry using the Sinistro cameras on Las Cumbres Observatory's network of 1\,m telescopes \citep{brown_cumbres_2013} as part of the Global Supernova Project. We subtracted reference images of the field taken with the same telescopes on 2022 May 27, about 4 months after the SN had faded, using PyZOGY \citep{guevel_pyzogy_2021} and measured PSF photometry on the difference images using \texttt{lcogtsnpipe} \citep{valenti_diversity_2016}. We calibrated this photometry to the AAVSO Photometry All-Sky Survey \citep{henden_aavso_2009}. $B$ and $V$ are reported in Vega magnitudes, and $g$, $r$, and $i$ are reported in AB magnitudes.

\begin{figure}
    \centering
    \includegraphics[width=\columnwidth]{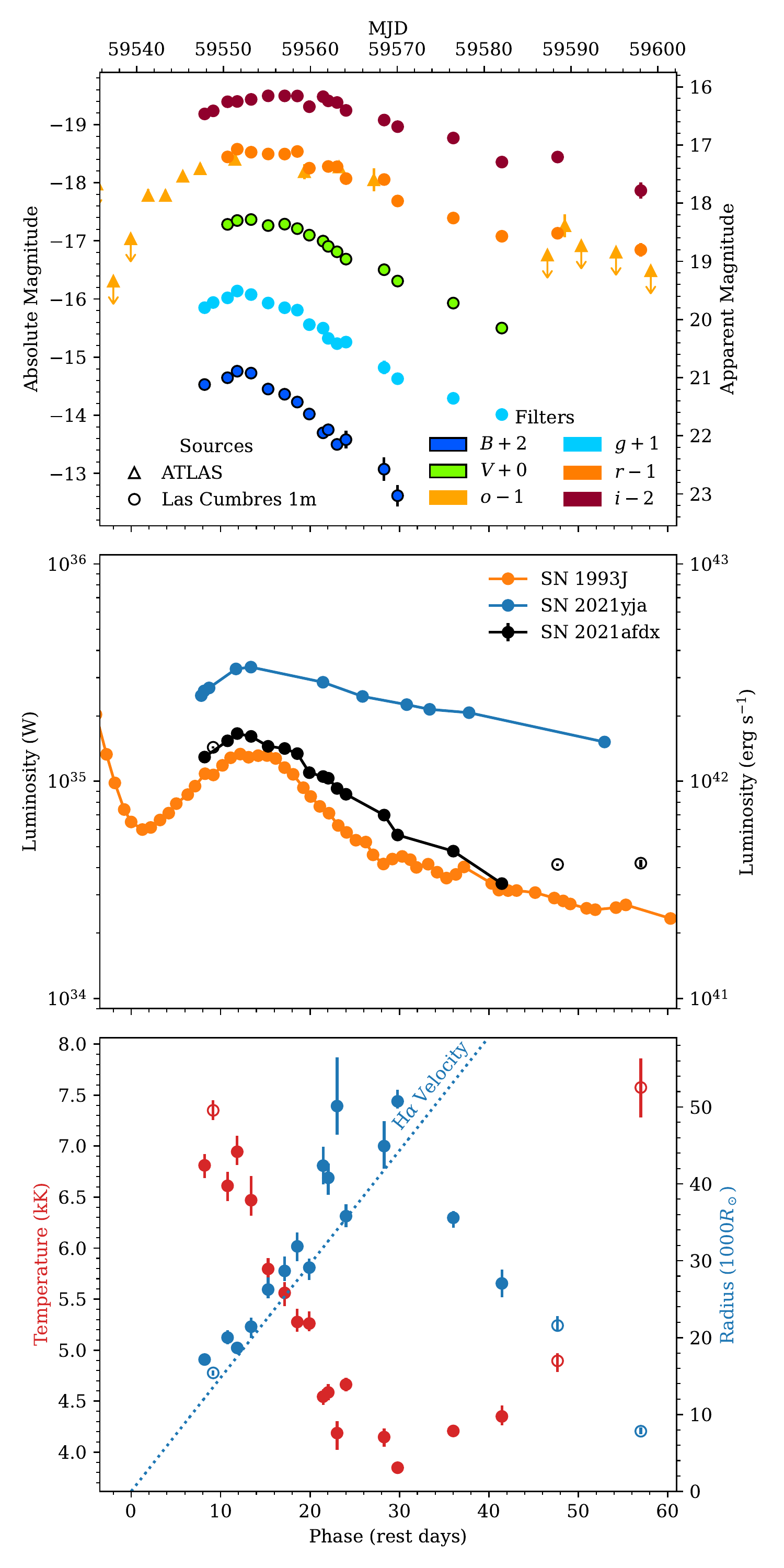}
    \caption{Top: ground-based optical light curve of SN~2021afdx from ATLAS and Las Cumbres Observatory, corrected for Milky Way extinction. Arrows indicate $3\sigma$ nondetections. Phases are given with respect to the last prediscovery nondetection. Center: pseudobolometric ($U$ to $I$) light curve of SN~2021afdx compared to the pseudobolometric light curves of the Type~IIb SN~1993J and the Type~II SN~2021yja, approximately aligned to their peak phase. Open markers indicate epochs with detections in only two filters, which may not yield reliable luminosity or temperature measurements. The gradual rise and decline of SN~2021afdx, rather than a rapid rise to a plateau, suggests a Type~IIb classification. Bottom: the photospheric temperature (red; left axis) and radius (blue; right axis) of SN~2021afdx derived from the data above. The dotted line shows that the ejecta radius inferred from the measured H$\alpha$ velocity is consistent with the photospheric radius around peak. (The data used to create this figure are available.)}
    \label{fig:phot}
\end{figure}

\defcitealias{planckcollaboration_planck_2020}{Planck Collaboration (2020)}
We adopt a luminosity distance of $d_\mathrm{L} = 136.8$~Mpc (distance modulus $\mu=35.68$~mag) based on the redshift of the Cartwheel Galaxy ($z=0.030187$; \citealt{amram_ha_1998}) and the cosmology of the \citetalias{planckcollaboration_planck_2020}. Absolute magnitudes are corrected for Milky Way extinction of $E(B-V)=0.0092$~mag \citep{schlafly_measuring_2011} using the \cite{fitzpatrick_correcting_1999} extinction law. Figure~\ref{fig:phot} (top) shows the ground-based optical light curve.

We construct a pseudobolometric light curve by fitting a blackbody SED to each epoch of photometry to get the photospheric temperature and radius, using a Markov Chain Monte Carlo (MCMC) routine implemented in the Light Curve Fitting package \citep{hosseinzadeh_light_2022}. We then integrate this SED from the $U$ to $I$ bands to obtain a pseudobolometric luminosity that is comparable to previous optical-only data sets. The results are shown in Figure~\ref{fig:phot} (center and bottom). The peak occurs at a phase of 11.8 days, with a pseudobolometric luminosity of $L_\mathrm{peak} = 1.66 \times 10^{42}\mathrm{\ erg\ s^{-1}}$ and a photospheric temperature of $T_\mathrm{peak} = 6900$~K.

\begin{figure*}
    \centering
    \includegraphics[width=\textwidth]{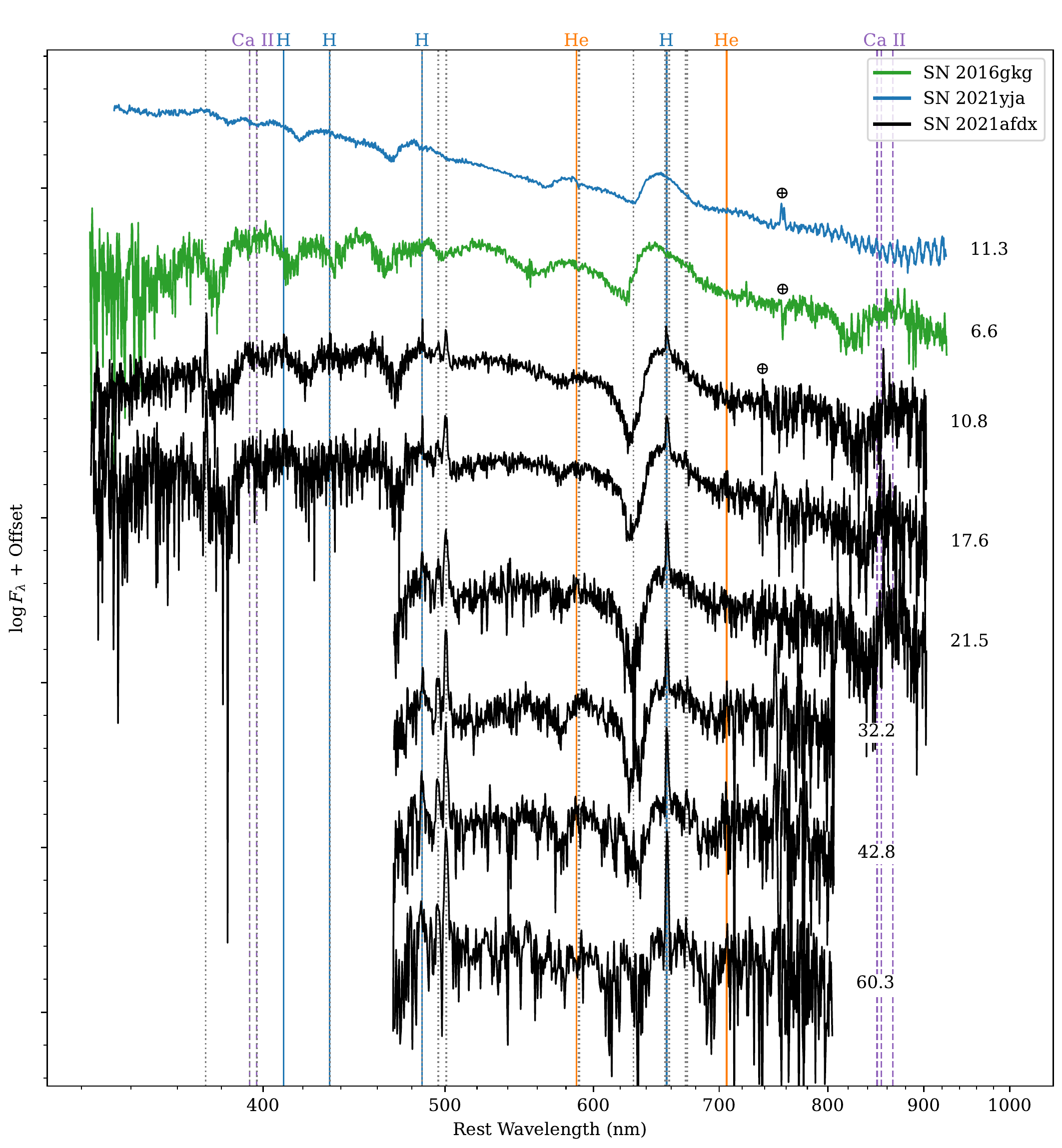}
    \caption{Spectral series of SN~2021afdx compared to spectra of the Type~IIb SN~2016gkg and the Type~II SN~2021yja, all from FLOYDS. Phases are given to the right of each spectrum in rest-frame days. The latest spectrum is binned by a factor of 2 for clarity. We attribute the narrow emission lines (marked with gray dotted lines) to the underlying star-forming region in the Cartwheel Galaxy. The strongest telluric feature is marked with the $\earth$ symbol. The remaining features can be explained by P~Cygni profiles of hydrogen, helium, and \ion{Ca}{2}. The broad, deep, asymmetric hydrogen absorption feature, the high helium-to-hydrogen line ratio, and relatively red continuum suggest a Type~IIb classification for SN~2021afdx, although we never observe the hydrogen feature fully disappear. (The data used to create this figure are available.)}
    \label{fig:spec}
\end{figure*}

SN~2021afdx was classified as an SN~II by the Advanced Extended Public ESO Spectroscopic Survey of Transient Objects (ePESSTO+; \citealt{smartt_pessto:_2015}) based on a spectrum taken with the ESO Faint Object Spectrograph and Camera 2 (EFOSC2; \citealt{buzzoni_eso_1984}) on the New Technology Telescope on 2021 November 26.173 \citep{ragosta_epessto+_2021}. We obtained six additional optical spectra using FLOYDS on Las Cumbres Observatory's 2\,m Faulkes Telescope South \citep{brown_cumbres_2013}, which are logged in Table~\ref{tab:spec} and plotted in Figure~\ref{fig:spec}. The spectra are available in machine-readable form in the online journal and on the Weizmann Interactive Supernova Data Repository \citep{yaron_wiserep_2012}.

\begin{deluxetable}{ccccC}
\tablecaption{Spectroscopic Observations and Velocities}\label{tab:spec}
\tablehead{\colhead{MJD} & \colhead{Telescope} & \colhead{Instrument} & \colhead{Phase} & \colhead{H$\alpha$ Velocity} \\[-6pt] &&& \colhead{(d)} & \colhead{(Mm s$^{-1}$)}}
\startdata
59550.492 & FTS & FLOYDS & 10.8 & 12.59 \pm 0.07 \\
59557.503 & FTS & FLOYDS & 17.6 & 11.70 \pm 0.07 \\
59561.491 & FTS & FLOYDS & 21.5 & 12.1 \pm 0.1 \\
59572.494 & FTS & FLOYDS & 32.2 & 11.9 \pm 0.1 \\
59583.469 & FTS & FLOYDS & 42.8 & 11.1 \pm 0.1 \\
59601.425 & FTS & FLOYDS & 60.3 & \nodata
\enddata
\end{deluxetable}
\vspace{-12pt}

Type~IIb SNe are a transitional class of core-collapse explosions in which early spectra show strong hydrogen lines that fade away in later spectra (see \citealt{gal-yam_observational_2016} for a review). These are thought to come from partially stripped yellow supergiant progenitors (see reviews by \citealt{smartt_progenitors_2009} and \citealt{van_dyk_supernova_2016}). \cite{ragosta_epessto+_2021a} raised the possibility of a Type~IIb subclassification for SN~2021afdx in their initial AstroNote, based on spectroscopic similarities to SN~2008aq. Several aspects of our data support this classification. First, the absorption component of the H$\alpha$ P~Cygni profile is very strong, broad, and asymmetrical compared to typical SN~II, and more closely resembles Type~IIb spectra. We show this in Figure~\ref{fig:spec}, where we compare to early spectra of the Type~IIb SN~2016gkg \citep{tartaglia_progenitor_2017} and the Type~II SN~2021yja \citep{hosseinzadeh_weak_2022}. The helium-to-hydrogen line ratio is also high and increasing in our spectral series, suggesting that the hydrogen features might have faded after the end of our observing campaign. Our earliest spectrum of SN~2021afdx is redder (Figure~\ref{fig:spec}) and the photospheric temperature is lower (Figure~\ref{fig:phot}, bottom) than typical SNe~II at this phase. Lastly, the gradual rise and decline of the bolometric light curve of SN~2021afdx more closely resembles the Type~IIb SN~1993J\footnote{We constructed the pseudobolometric light curve of SN~1993J using data from \cite{okyudo_v-band_1993}, \cite{vandriel_bvri_1993}, \cite{benson_light_1994}, \cite{lewis_optical_1994}, \cite{richmond_ubvri_1994,richmond_ubvri_1996}, \cite{barbon_sn_1995}, and \cite{metlova_observations_1995}.} than the Type~II SN~2021yja \citep{hosseinzadeh_weak_2022}, which rose quickly to a plateau (see Figure~\ref{fig:phot}, center). However, as we never see the hydrogen features fully disappear, even by day 60, and our light curve does not extend to late enough times to observe a potential fall from plateau (${\sim}100$ days), we cannot rule out a fast-declining SN~II (i.e., an SN~IIL) with some spectroscopic peculiarities. Our analysis does not depend strongly on the SN type, other than that it is the core collapse of a massive star, so we proceed by using the broader term (SN~II) and comparing to both SNe~II and IIb whenever possible.

We measure the photospheric velocity by fitting the sum of two equal-width Gaussians, one positive and one negative, and a linear continuum to the H$\alpha$ feature in each of the first five spectra. The signal-to-noise ratio in the last spectrum is too low to confidently fit the absorption component. We use an MCMC routine with uniform priors on the centers of the Gaussians and the continuum intercept and log-uniform priors on the amplitudes and width of the Gaussians and the continuum slope. We report the means and standard deviations of the velocity posteriors, as calculated from the maximum and minimum of the model minus the continuum, in Table~\ref{tab:spec}. The mean and standard deviation of these five measurements is $v_\mathrm{ej} \approx 11.9 \pm 0.5\mathrm{\ Mm\ s^{-1}}$. Figure~\ref{fig:phot} (bottom) shows that this velocity is consistent with the photospheric radii around peak.

\subsection{Space-based Infrared} \label{sec:obs_ir}

\begin{figure*}
    \centering
    \includegraphics[width=\textwidth]{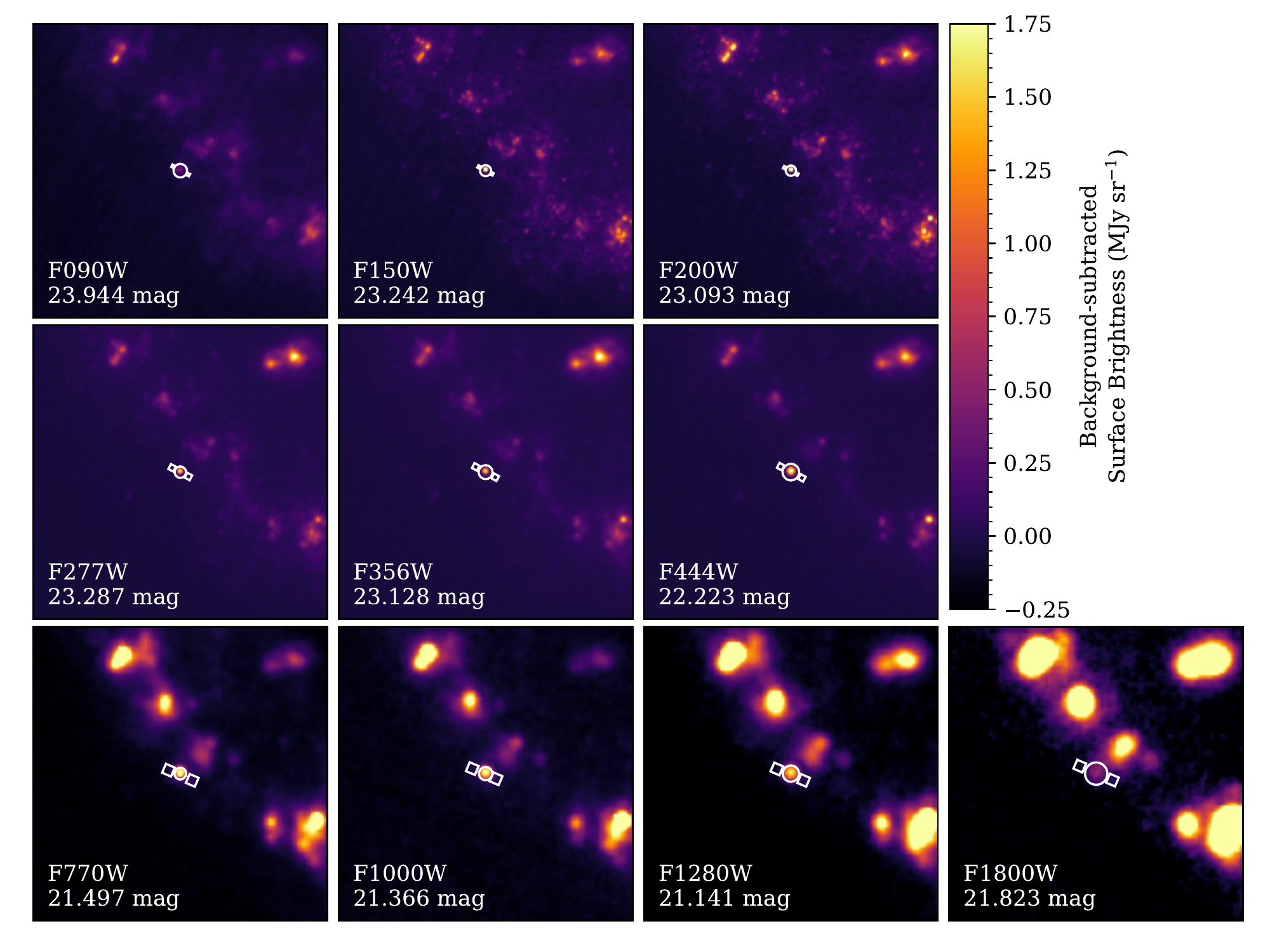}
    \caption{$10\arcsec\times10\arcsec$ cutouts of JWST images of the Cartwheel Galaxy centered on SN~2021afdx in 10 filters. The images are in their native resolution but have been resampled to have north up and east to the left. White circles mark the aperture used for photometry. White squares mark the region used for background subtraction. The filter name and SN brightness (in AB magnitudes) are reported in the lower left corner of each panel.}
    \label{fig:jwst}
\end{figure*}

We downloaded the JWST images of the Cartwheel Galaxy in 10 filters spanning 0.9--18~\micron{} from the Mikulski Archive for Space Telescopes (Proposal 2727; PI: Pontoppidan; doi:\dataset[10.17909/2n49-hx69]{https://doi.org/10.17909/2N49-HX69}) and performed aperture photometry using Photutils \citep{bradley_astropy_2022}. For NIRCam, we used a circular aperture containing 80\% of the PSF energy. For MIRI, we used a circular aperture containing 50\% of the PSF energy, to avoid contamination from the star-forming region. We then applied aperture corrections from the JWST Calibration Reference Data System \citep[CRDS;][]{greenfield_calibration_2016}. Table~\ref{tab:jwst} lists the results, given in AB magnitudes, and Figure~\ref{fig:jwst} shows cutouts of these 10 images centered on the SN.

\begin{deluxetable*}{ccccCCCc}
\tablecaption{JWST Observations and Photometry}\label{tab:jwst}
\tablehead{\colhead{MJD} & \colhead{Instrument} & \colhead{Filter} & \colhead{Exp.\ (s)\tablenotemark{a}} & \colhead{Magnitude} & \colhead{Flux ($\mu$Jy)} & \colhead{Bkg.\ Flux ($\mu$Jy)} & \colhead{Phase (day)}}
\startdata
59742.131 & NIRCam & F090W & 2748.616 & 23.944 \pm 0.030 & 0.960 \pm 0.018 & 1.553 \pm 0.021 & 197 \\
59742.172 & NIRCam & F150W & 2748.616 & 23.242 \pm 0.011 & 1.833 \pm 0.012 & 0.927 \pm 0.016 & 197 \\
59742.214 & NIRCam & F200W & 2748.616 & 23.093 \pm 0.007 & 2.103 \pm 0.011 & 0.713 \pm 0.012 & 197 \\
59742.131 & NIRCam & F277W & 2748.616 & 23.287 \pm 0.004 & 1.759 \pm 0.006 & 0.499 \pm 0.004 & 197 \\
59742.172 & NIRCam & F356W & 2748.616 & 23.128 \pm 0.009 & 2.036 \pm 0.015 & 0.696 \pm 0.017 & 197 \\
59742.214 & NIRCam & F444W & 2748.616 & 22.223 \pm 0.007 & 4.686 \pm 0.022 & 2.244 \pm 0.028 & 197 \\
59745.322 & MIRI & F770W & 4040.464 & 21.497 \pm 0.007 & 9.14 \pm 0.13 & 22.796 \pm 0.047 & 200 \\
59745.354 & MIRI & F1000W & 4040.464 & 21.366 \pm 0.008 & 10.32 \pm 0.11 & 71.842 \pm 0.037 & 200 \\
59745.388 & MIRI & F1280W & 4040.464 & 21.141 \pm 0.017 & 12.69 \pm 0.33 & 184.52 \pm 0.17 & 200 \\
59745.420 & MIRI & F1800W & 4084.864 & 21.823 \pm 0.076 & 6.77 \pm 0.60 & 1097.60 \pm 0.38 & 200 \\
\enddata
\tablenotetext{a}{Effective exposure time, corrected for dead time and lost time.}
\end{deluxetable*}
\vspace{-24pt}

The galaxy background in this region is complicated, and in the longer-wavelength filters the flux in the aperture is dominated by instrumental background. Thus, careful background subtraction is required. We chose two $3\times3$ pixel squares directly above and below the aperture (roughly perpendicular to the nearby star-forming region) and estimated the background level and its uncertainty by taking the median and the median absolute deviation, respectively, of those 18 pixels. Figure~\ref{fig:jwst} shows the source and background apertures in each image, and Table~\ref{tab:jwst} lists the source and background fluxes within the source aperture. We subtracted the background level from each pixel in the aperture and added the background uncertainty in quadrature to the error images provided. The choice of the background level may explain the discrepancy between our measurements and the preliminary photometry of \cite{engesser_detection_2022}.

At the time of our analysis, the MIRI aperture corrections had been updated using in-flight data (\texttt{jwst\_miri\_apcorr\_0008.fits}), but the NIRCam aperture corrections had not (\texttt{jwst\_nircam\_apcorr\_0004.fits}). In addition, the zero-point calibrations of both instruments suffer from uncertainties relative to preflight expectations. \cite{boyer_jwst_2022} report time-variable offsets of 1--23\% in the eight NIRCam detectors, and the most recent reduced MIRI images suffer from imperfect flat-fielding of up to ${\approx}5\%$ uncertainty in flux values. The data at ${>}18$~\micron{} has an additional uncertainty on flux zero-point, as their linearity correction coefficients have not yet been updated based on in-flight data. We choose to proceed with the official zero-points from CRDS (version 11.16.3), keeping in mind that our results can be reevaluated in the future when more accurate calibrations are established.

\section{Dust Modeling} \label{sec:dust}
\begin{figure*}[p]
    \centering
    \includegraphics[width=\columnwidth]{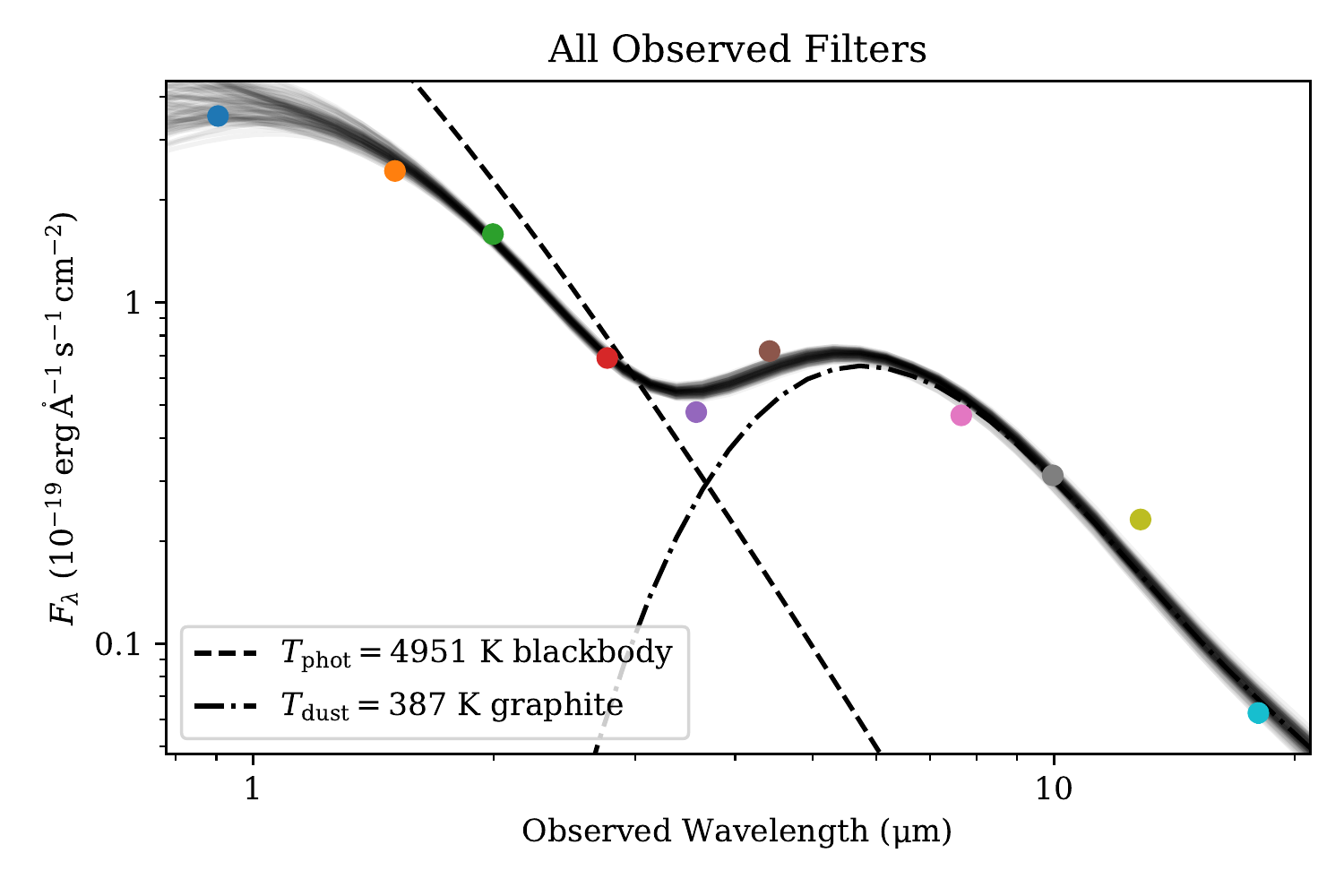}
    \includegraphics[width=\columnwidth]{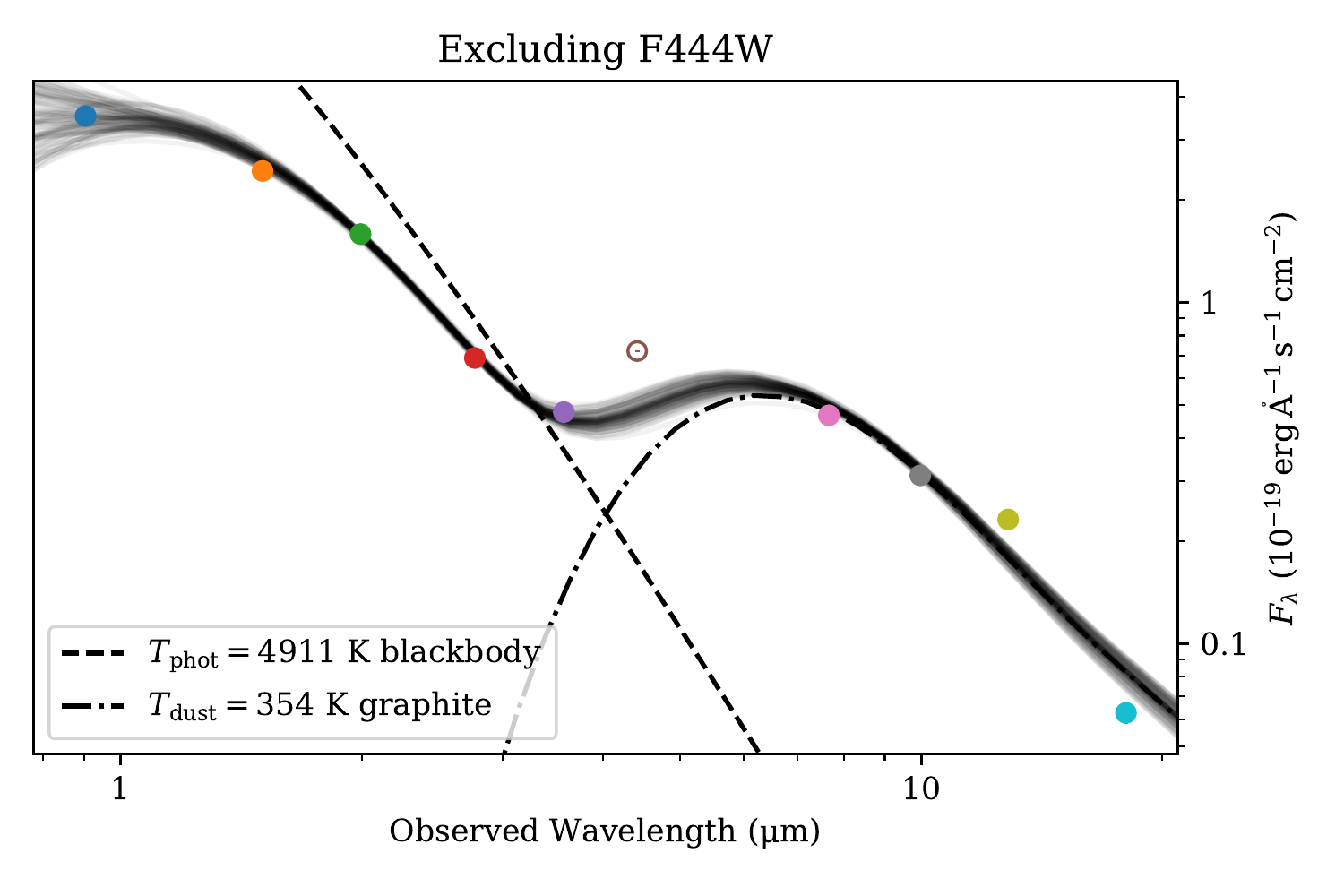}
    \includegraphics[width=\columnwidth]{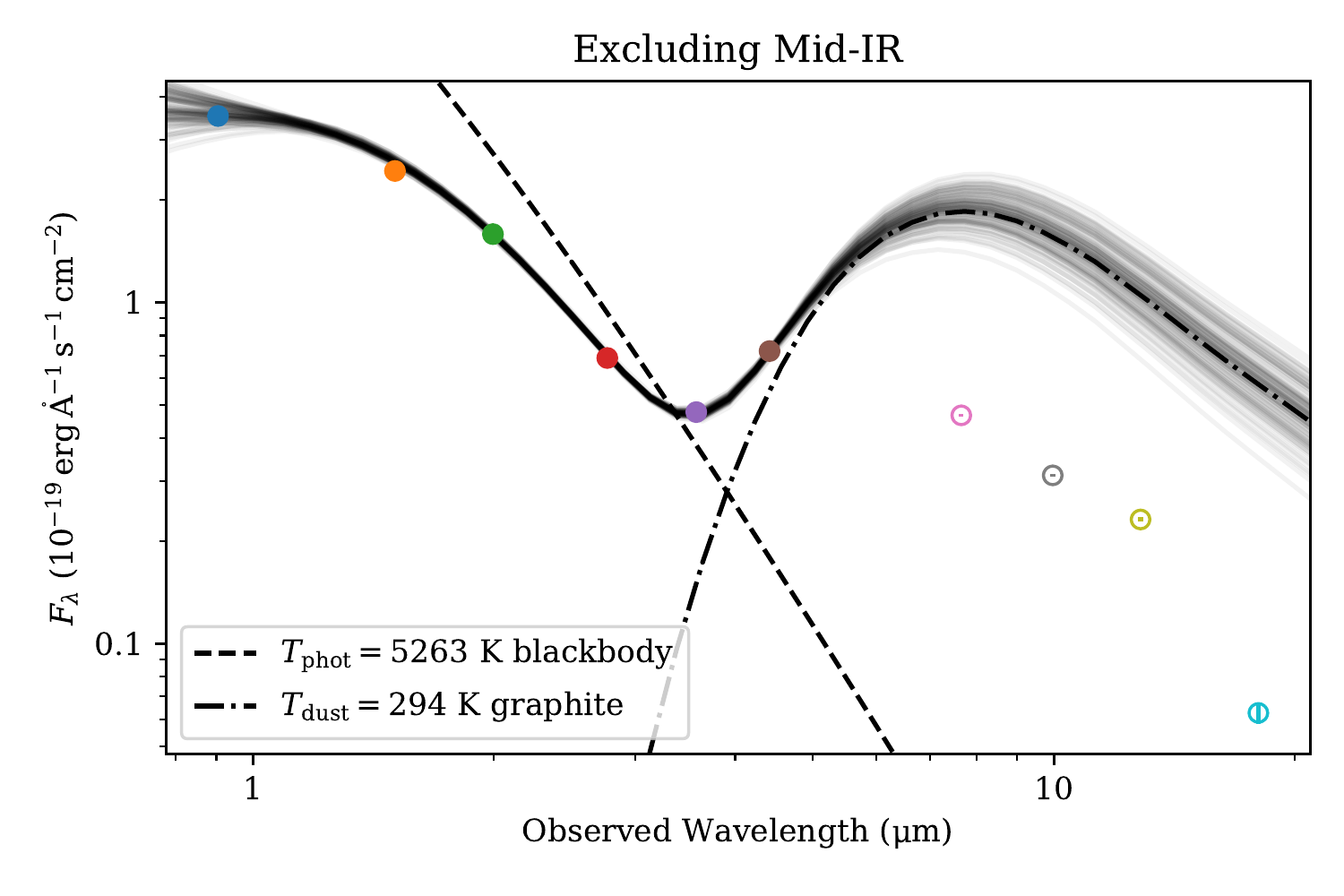}
    \includegraphics[width=\columnwidth]{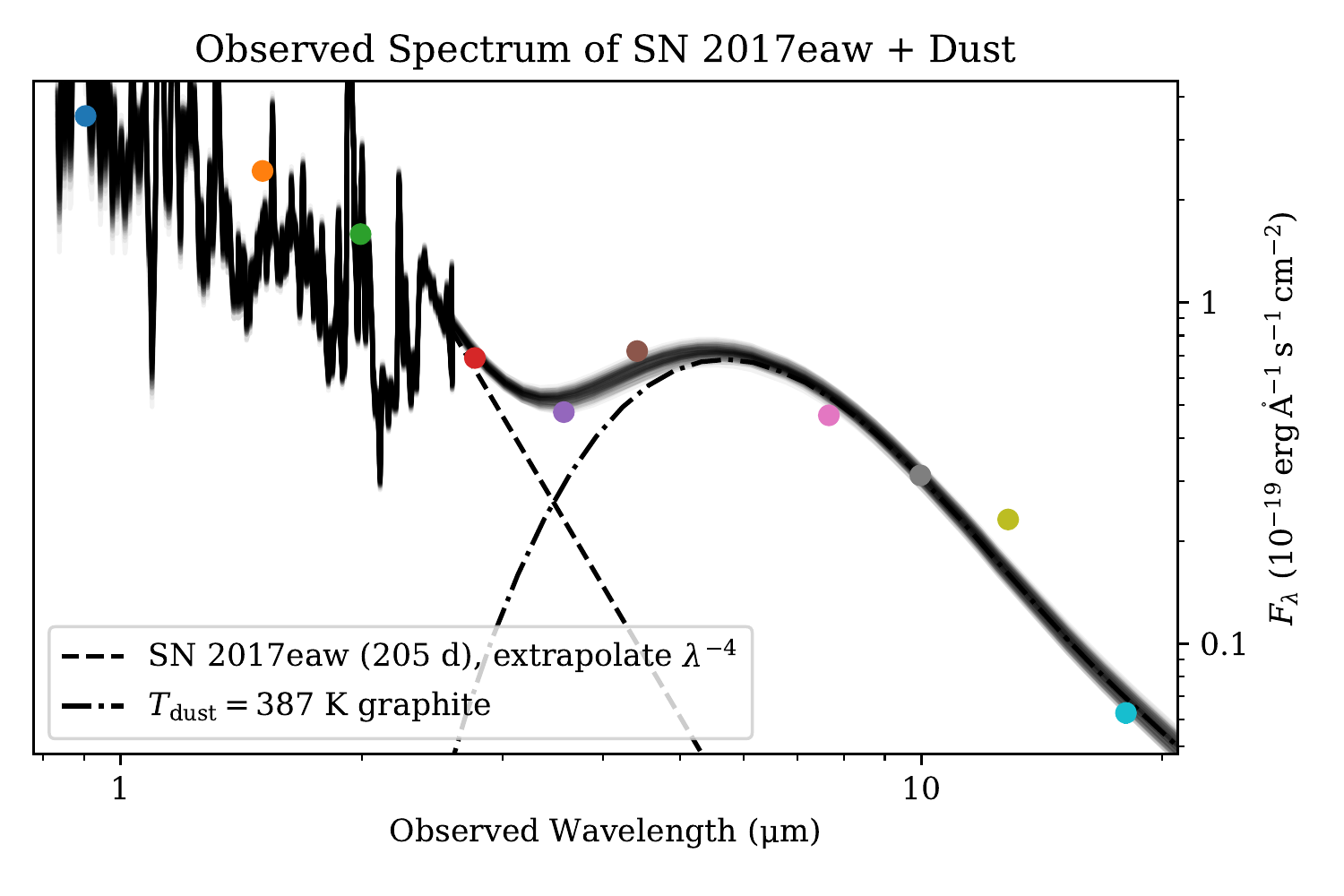}
    \caption{Top left: the IR SED of SN~2021afdx at a phase of 197--200 rest-frame days (circles), compared to models of dust emission (solid lines). The dashed line shows the hot blackbody component without extinction and the dotted--dashed line shows the warm graphite dust component. The dotted line shows a blackbody with the same temperature and radius as the dust. The blackbody plus dust model describes our observations well.
    Top right: same as left, but excluding F444W from the fit (open circle). This filter may be contaminated by carbon monoxide emission. Excluding it gives a slightly better fit to the remaining data points, so we adopt this as our preferred model.
    Bottom left: same as above, but excluding all filters redward of F444W (open circles), i.e., all the mid-IR observations. This wavelength coverage is analogous to previous SN observations with the Spitzer Warm Mission. The dust SED is very poorly constrained and is not consistent with our mid-IR observations, leading to incorrect inferences about the dust properties.
    Bottom right: same as others, but using the observed spectrum of SN~2017eaw \citep{rho_nearinfrared_2018}, extrapolated as $\lambda^{-4}$ (dashed line), to model the near-IR filters. The resulting dust parameters are nearly identical to the blackbody fit (top left) but with a slightly higher intrinsic scatter.}
    \label{fig:sed}
\end{figure*}

Figure~\ref{fig:sed} shows the infrared SED of SN~2021afdx at a phase of 197--200 rest-frame days. The most notable feature is a peak in $F_\lambda$ at 4--7~\micron, a second emission component that may indicate dust associated with the SN.

\begin{deluxetable*}{llCCCCc}[p]
\tablecaption{Dust Model Parameters} \label{tab:params}
\tablehead{&& \multicolumn{4}{c}{$1\sigma$ Equal-tailed Credible Interval} & \\[-10pt]
\colhead{Parameter} & \colhead{Prior} & \multicolumn{4}{c}{------------------------------------------------------------------------------} & \colhead{Unit} \\[-10pt]
&& \colhead{All Filters} & \colhead{No F444W} & \colhead{No Mid-IR} & \colhead{SN~2017eaw} &}
\startdata
Photospheric temperature & Uniform(0.1, 100) & 5^{+2}_{-1} & 5 \pm 1 & 5.3^{+0.7}_{-0.5} & \nodata & kK \\
Photospheric radius & Log-uniform(0.01, 1000) & 4.6^{+0.8}_{-0.7} & 4.9^{+1.0}_{-0.5} & 4.8 \pm 0.3 & \nodata & 1000 $R_\odot$ \\
Dust temperature & Uniform(0.1, 100) & 0.387^{+0.006}_{-0.005} & 0.354^{+0.006}_{-0.008} & 0.294^{+0.007}_{-0.006} & 0.387^{+0.005}_{-0.006} & kK \\
Dust radius & Log-uniform(0.01, 5000)\tablenotemark{a} & 1700^{+800}_{-400} & 1800^{+1200}_{-300} & 5700^{+1100}_{-800} & 5000^{+2000}_{-1000} & 1000 $R_\odot$ \\
Dust mass & Log-uniform(1, 1000) & 2.5 \pm 0.2 & 3.8^{+0.5}_{-0.3} & 40 \pm 10 & 2.5 \pm 0.2 & $10^{-3}\ M_\odot$ \\
Intrinsic scatter & Half-Gaussian(max=8)\tablenotemark{a} & 6.0 \pm 0.5 & 3.9^{+0.6}_{-0.4} & 2.1^{+0.6}_{-0.4} & 7.4 \pm 0.5 & \nodata \\
\enddata
\tablenotetext{a}{For the fit excluding the mid-IR filters, we extend the maximum on $R_\mathrm{dust}$ to $10^7\ R_\odot$ and reduce the prior on $\sigma$ to 6. \\For the fit with SN~2017eaw, we extend the maximum on $R_\mathrm{dust}$ to $10^7\ R_\odot$ and increase the prior on $\sigma$ to 10.}
\end{deluxetable*}
\vspace{-24pt}

We model the input luminosity as a hot blackbody with temperature $T_\mathrm{phot}$ and radius $R_\mathrm{phot}$ plus a uniform sphere of warm dust with temperature $T_\mathrm{dust}$, radius $R_\mathrm{dust}$, and mass $M_\mathrm{dust}$:
\begin{equation}\label{eq:Lin}
L_{\nu,\mathrm{in}} = 4 \pi R_\mathrm{phot}^2 \pi B_\nu(T_\mathrm{phot}) + 4 \kappa_\nu M_\mathrm{dust} \pi B_\nu(T_\mathrm{dust}),
\end{equation}
where $\kappa_\nu$ is the frequency-dependent opacity of the dust and $B_\nu(T)$ is the \cite{planck_vorlesungen_1906} function. We calculate the frequency-dependent opacity $\kappa_\nu$ from the absorption efficiency $Q_\nu$ and particle density $\rho_\mathrm{particle} = 2.26\mathrm{\ g\ cm^2}$ of $a = 0.1$~\micron{} graphite dust given by \cite{laor_spectroscopic_1993}:
\begin{equation}
    \kappa_\nu = \frac{3 Q_\nu}{4 a \rho_\mathrm{particle}} 
\end{equation}
Both components are extinguished according to the escape probability from \citet[Appendix~2]{osterbrock_astrophysics_1989}:
\begin{equation}
    p_\mathrm{esc} = \frac{3}{4\tau_\nu} \left[ 1 - \frac{1}{2\tau_\nu^2} + \left( \frac{1}{\tau_\nu} + \frac{1}{2\tau_\nu^2} \right) e^{-2\tau_\nu} \right],
\end{equation}
where the frequency-dependent optical depth to the center of the dust sphere with bulk density $\rho_\mathrm{bulk}$ is
\begin{equation}
\tau_\nu = \kappa_\nu \rho_\mathrm{bulk} R_\mathrm{dust} = \frac{3 \kappa_\nu M_\mathrm{dust}}{4 \pi R_\mathrm{dust}^2}.
\end{equation}
Therefore, the full model SED is
\begin{eqnarray}\label{eq:L}
    L_\nu = \left[ 4 \pi R_\mathrm{phot}^2 \pi B_\nu(T_\mathrm{phot}) \frac{3}{4 \tau_\nu} + 4 \pi R_\mathrm{dust}^2 \pi B_\nu(T_\mathrm{dust}) \right] \nonumber \\ \times \left[ 1 - \frac{1}{2\tau_\nu^2} + \left( \frac{1}{\tau_\nu} + \frac{1}{2\tau_\nu^2} \right) e^{-2\tau_\nu} \right].
\end{eqnarray}
In the optically thin limit ($\tau_\nu \ll 1$), Equation~\ref{eq:L} reduces to Equation~\ref{eq:Lin}, in which case $R_\mathrm{dust}$ will not be well constrained. In the optically thick limit ($\tau_\nu \gg 1$), the second term in square brackets in Equation~\ref{eq:L} approaches unity, in which case $M_\mathrm{dust}$ cannot be constrained independently of $R_\mathrm{phot}$.

We fit this model to the observed SED using an MCMC routine implemented in the Light Curve Fitting package \citep{hosseinzadeh_light_2022}.
We also include an intrinsic scatter term, $\sigma$, that accounts for uncertainties in the model (e.g., line emission) by inflating the error bars on each data point by a factor of $\sqrt{1 + \sigma^2}$. We run 20 walkers for 2000 steps to reach convergence and then another 1000 steps to sample the posterior. Table~\ref{tab:params} lists the model parameters, their priors, and their best-fit values (median and $1\sigma$ equal-tailed credible interval). The best-fit model, as well as a breakdown of the two components, is shown in Figure~\ref{fig:sed} (top left).

SNe~II (including SNe~IIb) are known to produce carbon monoxide (CO) during the nebular phase \citep[e.g.,][]{catchpole_spectroscopic_1988,spyromilio_carbon_1988,spyromilio_carbon_1996,ergon_type_2015}, whose fundamental rovibrational transition emits around 4.6~\micron. If CO is present in SN~2021afdx, this will produce an excess in our F444W observation with respect to the dust model. Therefore we repeat the above modeling procedure excluding this filter. Figure~\ref{fig:sed} (top right) and Table~\ref{tab:params} show the results. This indeed achieves a better fit ($\sigma = 3.9$ vs.\ 6.0) with a slightly lower dust temperature, undershooting F444W. We therefore adopt these results as our preferred set of parameters and claim a tentative detection of CO.

As a demonstration of the power of JWST in the mid-IR, we also repeat the fit excluding the four mid-IR filters (F770W, F1000W, F1280W, and F1800W). This fit simulates a data set from the Spitzer Warm Mission, the best IR data available over the past decade. Figure~\ref{fig:sed} (bottom left) and Table~\ref{tab:params} show that the dust SED is very poorly constrained without the mid-IR data points, and it leads to incorrect inferences about the dust properties.

\begin{figure}
    \centering
    \includegraphics[width=\columnwidth]{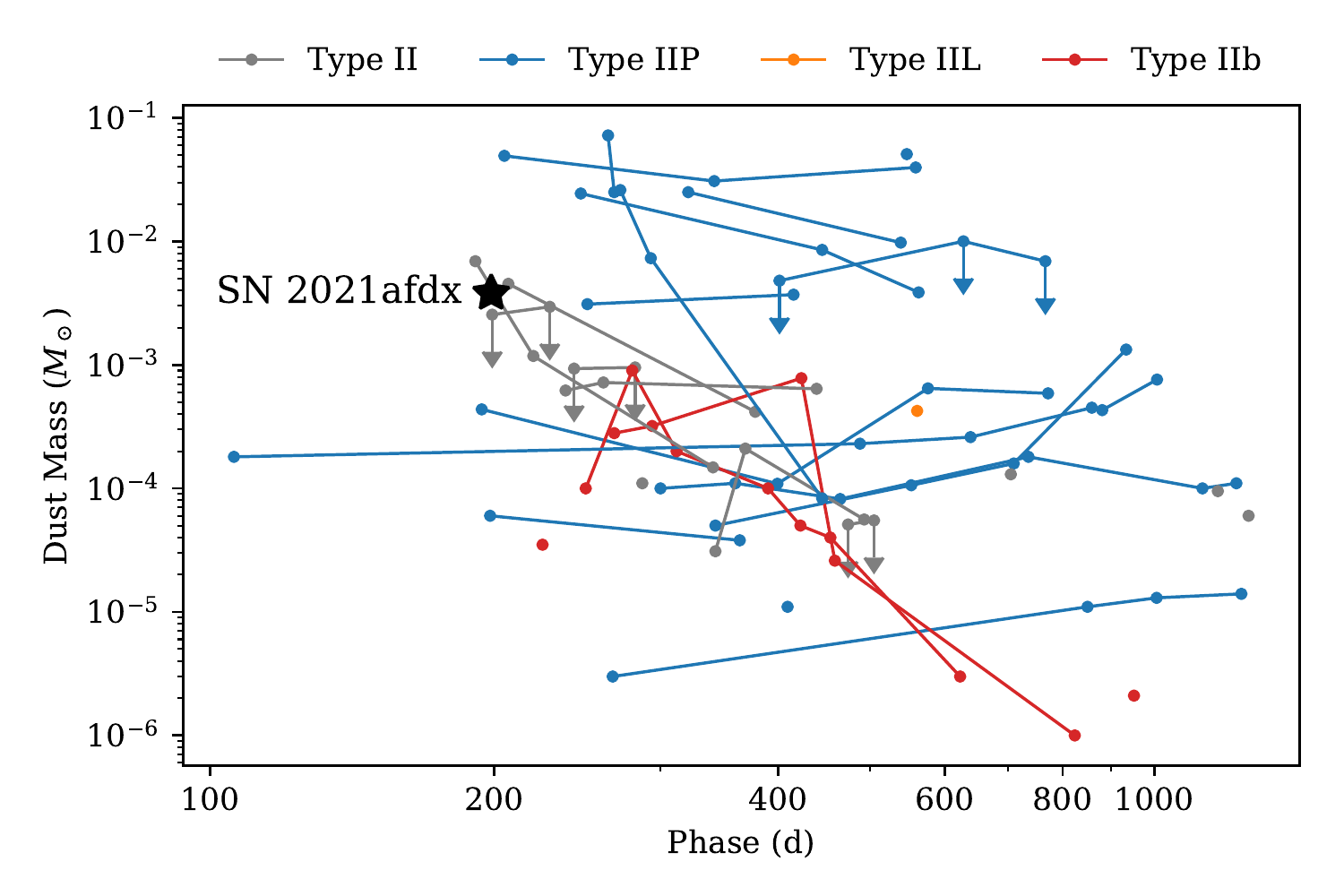}
\caption{Dust masses in SNe~II observed by Spitzer (circles) from \citet{szalai_comprehensive_2019}. Lines connect multiple measurements for individual SNe, and downward-pointing arrows indicate upper limits on dust masses when reference images for subtraction were not available. The dust mass of SN~2021afdx measured here (black star) is consistent with most of their distribution. Notably, our measurement is higher than the dust masses in any of the previously observed SNe~IIb, although the sample size is small.}
    \label{fig:dustmass}
\end{figure}

At ${\approx}200$ days, the near-IR SEDs of SNe may not be well approximated by a blackbody, due to strong nebular emission lines. Therefore, as a final check, we repeat the fit using a model in which the hot blackbody is replaced by the observed spectrum of the Type~II SN~2017eaw at 205 days after explosion from \cite{rho_nearinfrared_2018}, multiplied by a constant. We extrapolate from 2.52 to 19.5~\micron{} (although it is negligible at ${\gtrsim}5$~\micron) using a $\lambda^{-4}$ power law \citep{rayleigh_remarks_1900} matched to the red tail of the observed spectrum. The results, shown in Figure~\ref{fig:sed} (bottom right) and Table~\ref{tab:params}, give nearly identical dust parameters to the original fit with a blackbody, but with slightly higher intrinsic scatter.

Thanks to our simultaneous near-IR observations, which do not appear to suffer from significant extinction, we can conclude that the dust producing the mid-IR peak is optically thin. This implies a large radius for the dust sphere, on which our fitting procedure can only provide a lower limit, or deviations from our uniform spherical model (e.g., larger grains, clumps, or a nonspherical distribution). We discuss this further in Section~\ref{sec:discuss}. However, because the dust sphere is optically thin, the fit provides a true measurement of the dust mass: $M_\mathrm{dust} = (3.8_{-0.3}^{+0.5}) \times 10^{-3}\ M_\odot$.
This is consistent with most of the distribution of SN~II dust masses from \citet[see Figure~\ref{fig:dustmass}]{szalai_comprehensive_2019} and \cite{niculescu-duvaz_dust_2022}, although almost all of their observations are at later phases than ours. Notably, our limit is higher than the dust masses in any of the four SNe~IIb observed by \cite{szalai_comprehensive_2019}. The significance of this result will only become clear with larger sample sizes of the rarer subclasses of core-collapse SNe.

\section{Dust Formation Scenarios} \label{sec:discuss}
Given these measurements, we investigate whether the dust that we observe is newly formed in the SN ejecta or whether it existed in the circumstellar environment prior to explosion. Optical observations at this phase could show the effect of dust extinction on the broadband light curve or the spectroscopic line profiles, but at this late phase, SN~2021afdx is beyond the reach of most ground-based telescopes. Therefore we must rely only on the parameters from our IR SED modeling. In the newly formed dust scenario, we would expect the dust radius to be consistent with the ejecta radius and the dust temperature to be near the condensation temperature.

Using the mean H$\alpha$ velocities from Section~\ref{sec:obs_opt} and assuming homologous expansion, we estimate the ejecta position to be $R_\mathrm{ej}(t) = v_\mathrm{ej} t = (2.9 \pm 0.1) \times 10^5\ R_\odot$, which is smaller than the $3\sigma$ lower limit on the dust radius from Section~\ref{sec:dust}. Therefore the dust likely extends beyond, or lies entirely outside of, the SN ejecta. Theory predicts the condensation temperature for carbonaceous dust to be ${\gtrsim}1600$~K \citep{gall_production_2011}, about 4.5 times our measured dust temperature. However, astrophysical evidence from both massive stars \citep[e.g.,][]{beasor_evolution_2016,lau_revealing_2021} and SNe \citep[e.g.,][]{tinyanont_systematic_2016,szalai_comprehensive_2019} points to condensation temperatures below 1000~K, so we do not consider this a point against the newly formed dust scenario.

The alternative is that we are seeing an IR echo \citep[e.g.,][]{bode_infrared_1980,dwek_infrared_1983,graham_analysis_1986,meikle_spitzer_2006} off of preexisting dust in the circumstellar environment. We can check the consistency of this scenario with an equilibrium calculation: each dust grain must emit as much energy as it absorbs. The luminosity emitted by an individual dust grain with radius $a=0.1$~\micron{} and temperature $T_\mathrm{dust} = 354$~K is
\begin{equation}\label{eq:Lem}
L_\mathrm{em} = 4 \pi a^2 \pi \int B_\nu(T_\mathrm{dust}) Q_\nu d\nu.
\end{equation}
The luminosity absorbed by this dust grain from a source with photospheric radius $R_\mathrm{peak} = 18,700\ R_\odot$ and temperature $T_\mathrm{peak} = 6900$~K at a distance $R_\mathrm{dust}$ is
\begin{equation}\label{eq:Labs}
L_\mathrm{abs} = \pi a^2 \pi \int B_\nu(T_\mathrm{peak}) Q_\nu d\nu \frac{R_\mathrm{peak}^2}{R_\mathrm{dust}^2}.
\end{equation}
Setting Equation~\ref{eq:Lem} equal to Equation~\ref{eq:Labs}, we can solve for the radius of the dust shell.\footnote{Note that the result is not fully independent of $a$ because the efficiency curve $Q_\nu$ depends on the dust grain size.} We find that the dust must lie at $R_\mathrm{dust} = 3.5 \times 10^7\ R_\odot = 935$ light-days from the center of the explosion. This is fully consistent with the dust radius from Section~\ref{sec:dust}. Therefore we conclude that SN~2021afdx likely had dust in its circumstellar environment prior to explosion.

Although we cannot yet have observed the light reradiated from the entire sphere of dust due to light-travel time, we may be seeing emission only from the dust near the line of sight. If our interpretation is correct, further JWST observations of SN~2021afdx would show dust at this temperature for ${\approx}2000$ days before it begins to fade. However, this could also be a similar case to the Type~IIb SN~2011dh, which showed an IR echo from preexisting dust for the first ${\approx}100$ days but required newly formed dust or additional heating mechanisms to explain the IR light curve at ${\gtrsim}250$ days \citep{helou_mid-infrared_2013}.

\section{Summary and Conclusions} \label{sec:conclude}

We have presented near- and mid-IR observations of SN~2021afdx taken with JWST. The unprecedented combination of wavelength coverage (0.9--18~\micron) and sensitivity allows us to distinguish two distinct components in the nebular IR SED, which we attribute to hot ejecta and warm dust. By fitting models of dust emission to the SED, we measure the mass of dust to be $M_\mathrm{dust} = (3.8_{-0.3}^{+0.5}) \times 10^{-3}\ M_\odot$, which is fairly typical among core-collapse SNe at this phase and higher than all previously observed SNe~IIb dust masses.
We find that the radius of the dust sphere is significantly larger than the ejecta position at this phase, suggesting that we are seeing an IR echo off of preexisting dust in the progenitor environment. This means that SN~2021afdx has not (yet) produced dust in the amount needed to explain observations of galaxies in the early universe.

This Letter demonstrates the power of JWST to constrain models of SN dust formation, a capability that has been missing since the Spitzer Cold Mission. Furthermore, SN~2021afdx is almost twice as distant as the farthest SN~II observed by Spitzer in the nebular phase \citep{szalai_comprehensive_2019}, meaning that many more nebular SNe, including those of rarer subtypes (e.g., SNe~IIb), will be observable by JWST in the coming years. This increased sample size will quickly begin to probe the extent to which interstellar dust formation in the early universe can be explained by SNe.

\section*{Acknowledgments}

We thank Matteo Correnti for his JWST postpipeline data analysis notebook,\footnote{\url{https://jwst-docs.stsci.edu/jwst-post-pipeline-data-analysis/data-analysis-example-jupyter-notebooks}} which provided a starting point for our analysis, Samaporn Tinyanont for maintaining publicly available, machine-readable graphite opacity tables,\footnote{\url{https://github.com/stinyanont/sed_et_al}} and Jeonghee Rho for providing her spectrum of SN~2017eaw. We also thank Emma Beasor and Ryan Lau for helpful discussions about dust formation. Lastly, we thank the anonymous referee, whose careful consideration greatly improved this work.

This work is based on observations made with the NASA/ESA/CSA James Webb Space Telescope. The data were obtained from the Mikulski Archive for Space Telescopes at the Space Telescope Science Institute, which is operated by the Association of Universities for Research in Astronomy, Inc., under NASA contract NAS 5-03127 for JWST. These observations are associated with program \#2727.  The Early Release Observations and associated materials were developed, executed, and compiled by the ERO production team:  Hannah Braun, Claire Blome, Matthew Brown, Margaret Carruthers, Dan Coe, Joseph DePasquale, Nestor Espinoza, Macarena Garcia Marin, Karl Gordon, Alaina Henry, Leah Hustak, Andi James, Ann Jenkins, Anton Koekemoer, Stephanie LaMassa, David Law, Alexandra Lockwood, Amaya Moro-Martin, Susan Mullally, Alyssa Pagan, Dani Player, Klaus Pontoppidan, Charles Proffitt, Christine Pulliam, Leah Ramsay, Swara Ravindranath, Neill Reid, Massimo Robberto, Elena Sabbi, and Leonardo Ubeda. The EROs were also made possible by the foundational efforts and support from the JWST instruments, STScI planning and scheduling, and Data Management teams. The authors acknowledge the ERO production team for developing their observing program with a zero-exclusive-access period. This work makes use of data from the Las Cumbres Observatory telescope network.

Time domain research by G.H, D.J.S, and the University of Arizona team is supported by NSF grants AST-1821987, 1813466, 1908972, and 2108032, and by the Heising-Simons Foundation under grant \#2020-1864. 
J.E.A.\ is supported by the international Gemini Observatory, a program of NSF's NOIRLab, which is managed by the Association of Universities for Research in Astronomy (AURA) under a cooperative agreement with the National Science Foundation, on behalf of the Gemini partnership of Argentina, Brazil, Canada, Chile, the Republic of Korea, and the United States of America.
Research by S.V. is supported by NSF grant AST-2008108.
T.S. is supported by the NKFIH/OTKA FK-134432 grant of the National Research, Development and Innovation (NRDI) Office of Hungary, the J\'anos Bolyai Research Scholarship of the Hungarian Academy of Sciences, and the New National Excellence Program (\'UNKP-21-5) of the Ministry for Innovation and Technology of Hungary from the source of NRDI Fund. The Las Cumbres Observatory group is supported by NSF grants AST-1911225 and 1911151.

\facilities{ADS, JWST (MIRI, NIRCam), LCOGT (FLOYDS, Sinistro), OSC, NED, TNS.}

\defcitealias{astropycollaboration_astropy_2022}{Astropy Collaboration 2022}
\software{Astropy \citepalias{astropycollaboration_astropy_2022}, emcee \citep{foreman-mackey_emcee_2013}, FLOYDS pipeline \citep{valenti_first_2014}, \texttt{lcogtsnpipe} \citep{valenti_diversity_2016}, Light Curve Fitting \citep{hosseinzadeh_light_2022}, JWST CRDS \citep{greenfield_calibration_2016}, Photutils \citep{bradley_astropy_2022}, PyZOGY \citep{guevel_pyzogy_2021}.}

\bibliography{zotero_abbrev}

\end{document}